# Knowledge-data fusion framework for frequency security assessment in low-inertia power systems

## Author Information

**Yurun Zhang[1†], Wei Yao[1*†], Yutian Lan[1], Hang Shuai[1], Wei Gan[2], Shanyang Wei[1], Chao Duan[3*], Jinyu Wen[1*], Shijie Cheng[1]**

1 State Key Laboratory of Advanced Electromagnetic Technology, School of Electrical and Electronic Engineering, Huazhong University of Science and Technology, Wuhan, 430074, China.

2 School of Engineering, Cardiff University, Cardiff, CF24 3AA, United Kingdom.

3 School of Electrical Engineering, Xi'an Jiaotong University, Xi'an, 710049, China.

† These authors contributed equally

## Abstract

The integration of renewable energy via power electronics is transforming power grids into low-inertia systems, heightening the risks of frequency insecurity and widespread outages. Therefore, frequency security assessment (FSA) methods are urgently needed to ensure the reliable system operation. Recently, knowledge-data fusion models attempt to address the limitations of knowledge-driven (accuracy) and data-driven (generalization) FSA methods. However, current methods remain confined to shallow knowledge-data integration due to challenges in representing heterogeneous knowledge and establishing interactive mechanisms. Here, by classifing FSA domain knowledge into physics-guided and physics-constrained categories, we propose a guided learning-constrained network (GL-CN) framework, which deeply integrates domain knowledge across both network architecture and training process. In this framework, a data-driven model with dual input channels combining graph convolutional networks (GCN) and multilayer perceptrons (MLP) is proposed to extract both nodal and system-level power system features. Furthermore, guided learning enhances model generalization through data augmentation in pre-training utilizing physics-guided knowledge, while constrained network encodes physics-constrained knowledge into the network architecture and loss function to ensure physics-consistent and robust predictions. Validated on Yunnan Provincial Power Grid in China, our method reduces FSA time from days to seconds compared to traditional simulation, achieving 98% accuracy, robustness against 39.0% knowledge error, and generalization for 40%-60% renewable penetration. This provides a solid solution for mitigating blackouts caused by frequency insecurity and offers a generalizable paradigm for broader cross-domain problems.

## 1 Introduction

Power systems are among the most complex engineered infrastructures, requiring frequency to be maintained within strict operational limits to ensure security, known as frequency security[1]. Deviations beyond these limits (e.g., ±0.5Hz) can trigger catastrophic blackouts, resulting in substantial economic losses[2]. However, renewable integration via power electronics reduces system rotational inertia (inherent in synchronous generators) and consequently leads to low-inertia power systems[3-6]. In low-inertia power systems, frequency security is a predominant concern, as contingencies like renewable generation tripping and high-voltage direct current (HVDC) blocking can trigger severe frequency deviations and complex dynamic responses[7-10]. Recent large-scale blackouts including the 2016 South Australia blackout[11] and the 2019 Great Britain outage[12] highlight this concern. In South Australia (48.36% renewable generation penetration), severe weather caused widespread disconnection of renewable generation, resulting in a frequency caused blackout that left 1.7 million people without power. In Great Britain (30% renewable generation penetration), a lightning strike triggered initial frequency dips that cascaded into further wind turbine trips, causing severe underfrequency and outages affecting over 1 million customers. Given the growing threat to frequency security in low-inertia power systems, an accurate and effective frequency security assessment (FSA) method is urgently needed to mitigate such frequency insecurity events[13].

Power system frequency security relies on real-time balance between generation and load[14-15]. Under pre-fault steady-state conditions, generation precisely matches demand, maintaining a near-constant system frequency[16]. Following a fault, the power system enters a transient state with a generation-load power imbalance (denoted as $\Delta P$), resulting in frequency deviation (denoted as $\Delta f$). The system deals with this deviation through two fast frequency response mechanisms, including the inertial response of rotating masses (approximately 0–10 seconds, denoted as the inertia $H$)[17-18] and primary frequency control from governors (approximately 2–30 seconds, characterized by the regulation coefficient $R$)[19-20]. These mechanisms are designed to handle the frequency deviation for the majority of faults. However, a subset of severe faults may exceed the regulation capacity of power systems. Such faults require operator intervention through preventive or emergency control, constituting the central focus of FSA.

From a temporal perspective, FSA is typically categorized into pre-fault FSA conducted in advance to inform preventive control and post-fault FSA performed after a fault to support emergency control actions. Post-fault FSA supports emergency control strategies by predicting the full transient frequency response using pre-fault steady-state conditions and immediate post-fault measurements[21]. However, inherent delays in post-fault data collection render post-fault FSA and its associated emergency control reactive, potentially expanding fault impacts[22]. Therefore, proactive pre-fault FSA is increasingly prioritized to enable preventive control. This proactive paradigm aims to eliminate risks by adjusting the operating point[23-25] or mitigate losses by developing strategies in advance for immediate deployment upon fault occurrence[26]. Therefore, this work focuses on pre-fault FSA as the basis for preventive control. This method periodically acquires the steady-state operating point against a pre-defined set of anticipated faults[27]. For each fault, it evaluates key frequency metrics, including maximum rate of change of frequency ($RoCoF_{max}$), frequency nadir ($f_{nadir}$), time of frequency nadir ($t_{nadir}$), and quasi-steady-state frequency ($f_{ss}$), and thereby assesses the security risk under current operating conditions. Unlike post-fault FSA, which prioritizes computational speed, pre-fault FSA emphasizes accurate prediction of frequency metrics and model robustness and generalization in the absence of post-fault transient data.

Methodologically, FSA approaches are categorized as knowledge-driven (KD) or data-driven (DD) paradigms, each with distinct advantages and limitations. The KD methods, including precise time-domain simulation (TDS) models[28-29] and simplified aggregated system frequency response (ASFR) models[30-32], employ power system mechanisms for modeling, which gives them the capability to adapt to various operating conditions. However, in large-scale low-inertia power systems, precise TDS models require prohibitive computational costs (days for all the operating conditions), while simplified ASFR models introduce significant systematic errors due to idealized assumptions. These systematic errors of ASFR models introduce substantial frequency security risks, which are unacceptable for power system operators. This forces the grid to rely on time-consuming TDS methods for annual validation, with updates for new operating conditions remaining impractically time-consuming. In contrast, the DD methods establish a direct nonlinear mapping between power system states and frequency security risks, typically via deep learning models and data training[33]. The exceptional nonlinear mapping capabilities and computational speed offer a potential solution to the limitations of KD methods[34]. However, the inherent drawbacks of the DD methods raise significant concerns about the reliability of their predictions[35]: (1) Data dependence: the accuracy of DD models is significantly impacted by the quality and quantity of available training data[36]; (2) Generalization: performance degrades severely (with approximately a 10% decrease reported in Ref. 37) when renewable penetration is significantly different from that in the training data[38]. Therefore, the DD methods have not been successfully deployed in practical power systems due to the inability to adapt grid topology or operating condition changes.

Recent research in knowledge-data fusion represents a preliminary exploration to leverage the complementary strengths of KD and DD methods, demonstrating initial progress in power system FSA[39-40]. Ref. 41 adopted a serial architecture where the ASFR model outputs served as input features to a DD model, which then refined these preliminary predictions to obtain accurate post-fault frequency metrics. Ref. 42 optimized the transfer function of the KD model by comparing its output with that of a DD model, thereby obtaining an accurate post-fault frequency trajectory. While these studies demonstrate that knowledge-data fusion improves model accuracy, they remain limited in enhancing robustness and generalization. This limitation stems from the fact that existing methods are constrained to shallow fusion, primarily due to the absence of a framework for representing and integrating heterogeneous knowledge forms. Consequently, current work often approximates domain knowledge as simplified physical models, integrating them with DD models via basic serial or parallel connections. Although this may improve accuracy, it fails to overcome the inherent generalization deficit of DD approaches. More critically, the lack of an in-depth interactive mechanism between domain knowledge and DD models leaves the fusion models vulnerable to errors from both DD and KD modules, which in turn compromises model robustness and leaving it prone to unexpected errors in real-world scenarios with limited data or imprecise knowledge. Therefore, it's necessary to establish a knowledge-data fusion framework that effectively represents and integrates heterogeneous domain knowledge and establish a dedicated bidirectional interactive mechanism for the interconnection of knowledge and data.

In this work, we develop a novel knowledge-data fusion framework with knowledge representation and interaction to enhance the accuracy, robustness, and generalization of pre-fault FSA in low-inertia power systems. In this framework, domain knowledge is categorized into physics-guided knowledge, represented as approximate input-output mappings to guide the neural network pre-training process, and physics-constrained knowledge, expressed as inequality constraints between model outputs to impose theoretical bounds on predictions. A dual-channel neural network combining multilayer perceptrons (MLP) and graph convolutional networks (GCN) is designed to extract both nodal and system-level input features. To realize the deep integration of domain knowledge with DD models, we introduce a guided learning-constrained network (GL-CN) algorithm. **Guided learning (GL) incorporates physics-guided knowledge via data augmentation during pre-training to enhance model generalization and reduce data dependency. The constrained network (CN) integrates physics-constrained knowledge into the output layer structure and a knowledge constraint loss function, ensuring predictions remain within theoretical bounds.** Case studies conducted on Yunnan Provincial Power Grid (3,734 nodes), which is a typical low-inertia power system in China with 48.6% renewable generation penetration, demonstrate superior performance: (1) **Accuracy:** achieves an average 2.24% prediction error in frequency metrics, representing a 77.7% and 47.7% reduction over KD and DD methods; (2) **Robustness:** maintains <5% error using only 951 samples, outperforming DD methods by a 75% error reduction (data-limited scenarios), and maintains <3% error toleratint 39.0% knowledge error (knowledge-imprecise scenarios); (3) **Generalization:** maintains 97.96% FSA accuracy across unseen renewable penetration scenarios, representing a 4.60% and 3.17% accuracy improvement over KD and DD methods. Our framework overcomes the key barriers to deploying artificial intelligence (AI) methods in low-inertia power systems and reduces the FSA time from days to seconds, dramatically improving the efficiency of preventive control. Furthermore, the methodological paradigm is generalizable and holds potential for broader AI applications in other engineering fields such as weather forecasting and battery management.

# 2 Results

## 2.1 Knowledge-data fusion framework for pre-fault FSA

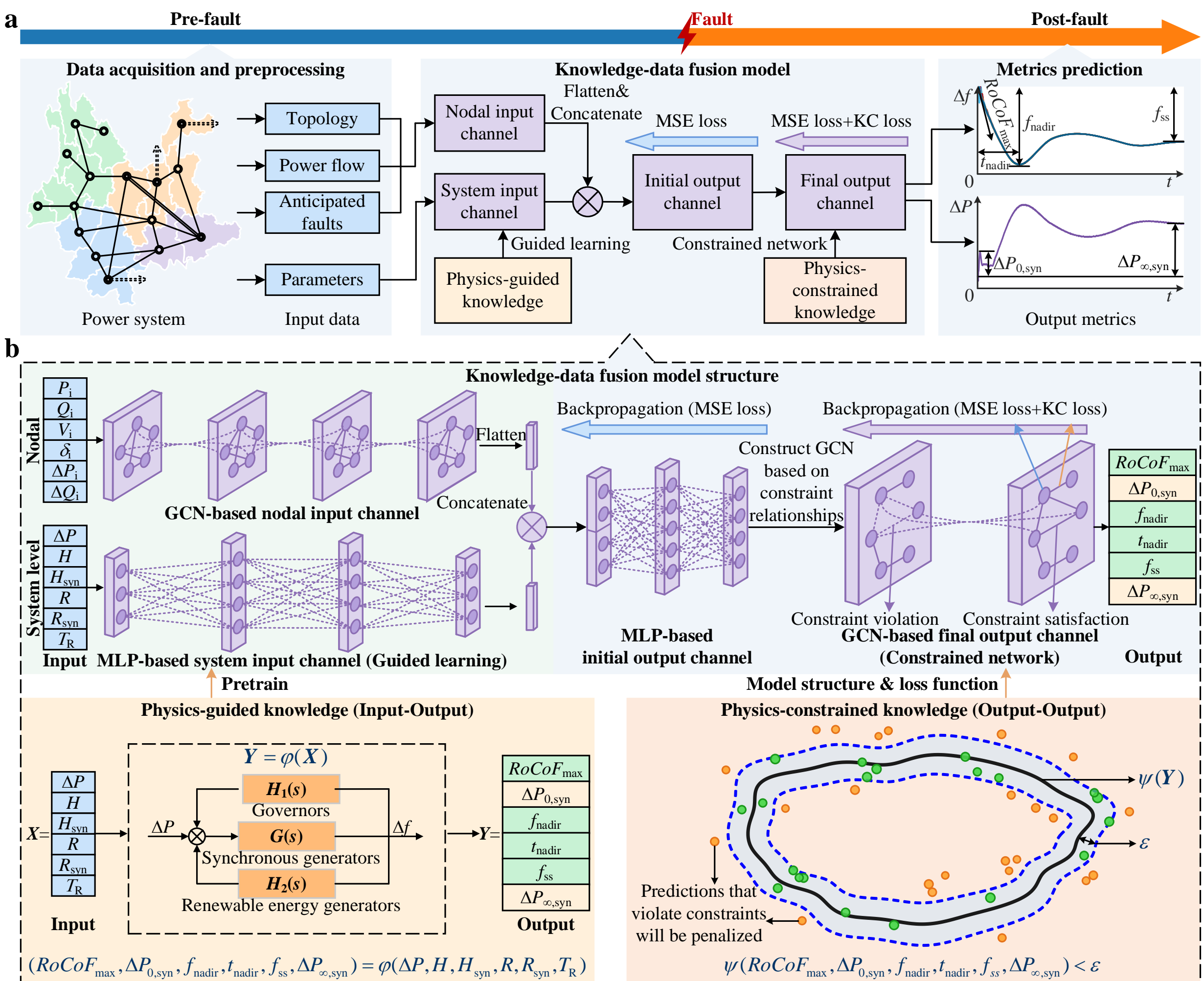

**Fig. 1: Proposed knowledge-data fusion framework for pre-fault FSA in low-inertia power systems.** **a Proposed framework for pre-fault FSA.** Four types of steady-state data are acquired: (1) topology, (2) power flow, (3) anticipated faults, and (4) frequency control parameters. The first three categories represent the nodal information, whereas the frequency control parameters describes the system-level information. The two categories of data are fed into the dual channels of the knowledge-data fusion model to yield the key predictive metrics for FSA. Beyond the four primary frequency metrics ($RoCoF_{max}$, $f_{nadir}$, $t_{nadir}$, and $f_{ss}$), the model also outputs power metrics, including initial power transient ($\Delta P_{0,\mathrm{syn}}$) and quasi-steady-state power adjustment ( $\Delta P_{\infty,\mathrm{syn}}$ ) of synchronous generators. **b Architecture of the proposed knowledge-data fusion model.** Two parallel input channels independently extract nodal and system-level features. These features are flattened, concatenated, and processed by a MLP to generate an initial prediction. This preliminary output is subsequently refined by a GCN that encodes physics-constrained knowledge to produce the final result. Within this framework, the physics-guided knowledge is modelled as the input-output mapping and embedded within the MLP-based system input channel. The physics-constrained knowledge, formulated as inequality constraints, is incorporated via the GCN-based final output channel.

The pre-fault FSA framework consists of three components: power system data acquisition and preprocessing (left), knowledge-data fusion model (middle), and frequency metrics prediction (right), as shown in **Fig. 1a**. During steady-state operation, the system acquires (1) topology as an adjacency matrix and (2) steady-state power flow (active power $P_i$, reactive power $Q_i$, voltage $V_i$, and phase angle $\delta_i$ of each node) from the low-inertia power system at a 30-second interval. Simultaneously, the (3) operator-defined set of anticipated faults (active power imbalance $\Delta P_i$, reactive power imbalance $\Delta Q_i$) is combined with steady-state operation data to create a feature matrix. Then, we assess (4) frequency control parameters (besides $H$, $R$, there are also inertia of synchronous generators $H_{\mathrm{syn}}$, regulation coefficient of synchronous generators $D_{\mathrm{syn}}$, and equivalent time constant of primary frequency regulation $T_{\mathrm{R}}$) under current operating conditions. The input features are categorized into nodal and system-level dimensions based on the information they represent, and are respectively fed into the dual channels of the knowledge-data fusion model. The model outputs characterize the system's frequency dynamics ($RoCoF_{\mathrm{max}}$, $f_{\mathrm{nadir}}$, $t_{\mathrm{nadir}}$, $f_{\mathrm{ss}}$) and power dynamics (initial power transient $\Delta P_{0,\mathrm{syn}}$, quasi-steady-state power adjustment $\Delta P_{\infty,\mathrm{syn}}$ of synchronous generators) following the fault. Formal definitions of these metrics are provided in **Methods**. We focus exclusively on synchronous machine dynamics to represent power dynamics, as their frequency-power relationships are both explicit and accurate. A system is considered to exceed acceptable frequency security limits when any of the following thresholds is violated: $RoCoF_{\mathrm{max}}$, $f_{\mathrm{nadir}}$ and $f_{\mathrm{ss}}$. Such threshold violations indicate substantial frequency deviation and elevated risk of frequency insecurity.

To integrate domain knowledge of low-inertia power systems with deep learning models, we designed an architecture consisting of three core components: input channel (green background), output channel (blue background), and knowledge module (orange background), as shown in **Fig. 1b**. Within the input channel, nodal data are encoded as graph-structured representations for subsequent processing by the GCN channel, while system-level data are fed into the MLP channel. In the output channel, high-level features extracted by the model are first transformed into initial outputs by a simple MLP, which are subsequently refined by a GCN, incorporating physics-constrained knowledge to generate the final outputs. In the knowledge module, we capture the mapping between frequency control parameters and security metrics based on ASFR model. The derivation is provided in **Supplementary Note 1**. This physics-guided relationship, which connects system inputs to outputs, is embedded within the MLP channel to improve the generalization of the model under varying parameters and reduce reliance on real-world data. In addition, our preliminary analysis revealed that the shape of the frequency response curve and the power imbalance distribution principle after faults impose constraints among the output metrics, which is detailed in **Methods**. We therefore embedded the physics-constrained knowledge within the final output channel to regularize the results and enhance the robustness of the model.

## 2.2 GL-CN algorithm for knowledge-data fusion

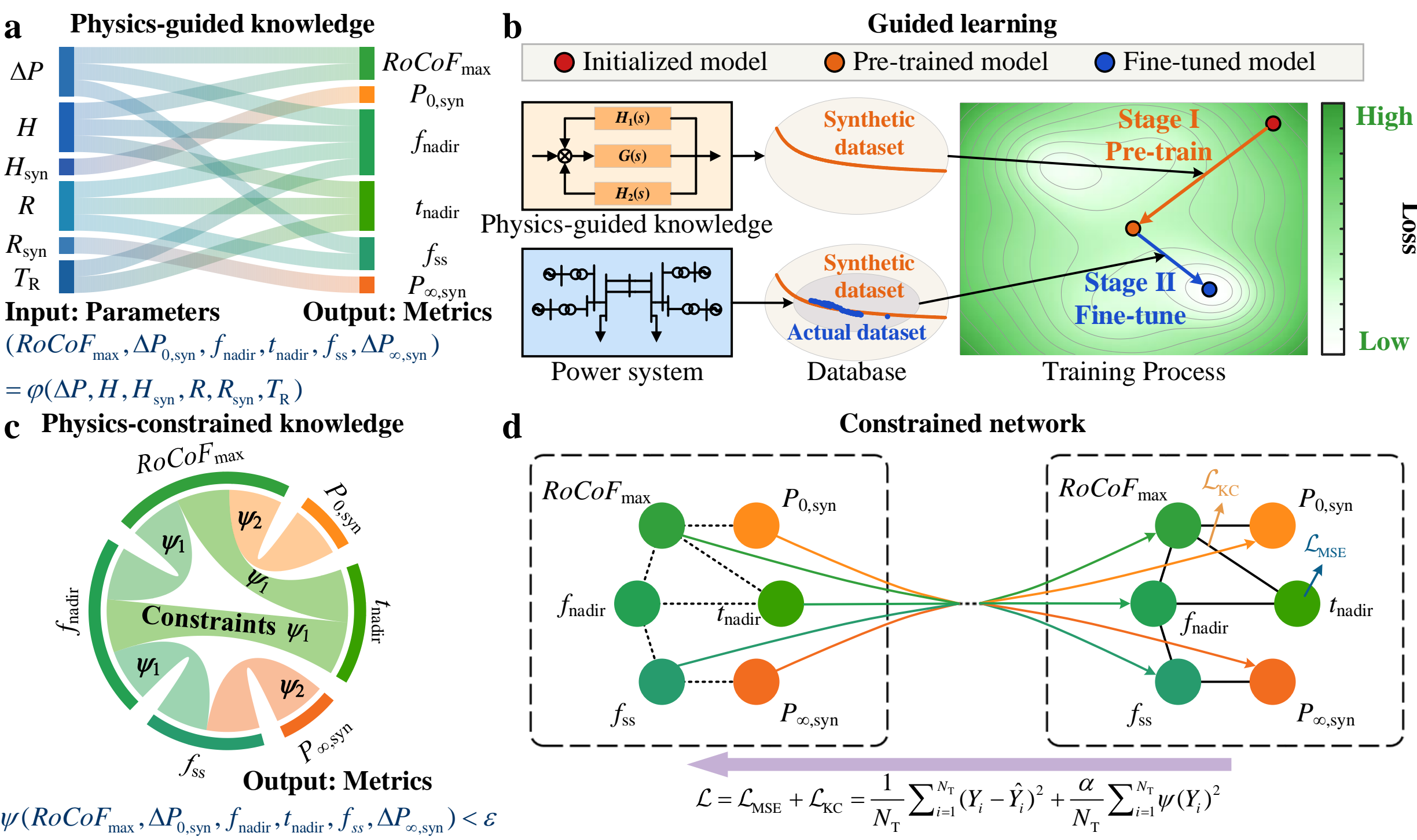

**Fig. 2: GL-CN algorithm for the construction and training of knowledge-data fusion model.**

**a Schematic of physics-guided knowledge.** The physics-guided knowledge encapsulates the quantitative relationships between system-level parameters and output metrics. The connecting bands represent a dominant influence of the linked input parameters on the corresponding output metric. $\varphi(\cdot)$ denotes the mapping relationship from inputs to outputs. **b Schematic of the proposed GL algorithm.** The algorithm proceeds in two stages. In the pre-training stage (Stage I), a synthetic dataset is generated by sampling from the physics-guided knowledge across a parameter space broader than actual operating conditions. The model is pre-trained on this dataset to learn the underlying physical principles. In the fine-tuning stage (Stage II), the model is refined and optimized using the actual dataset acquired from the power system. **c Schematic of physics-constrained knowledge.** This knowledge formalizes inequality relationships governing the output metrics ($RoCoF_{max}$, $f_{nadir}$, $t_{nadir}$, $f_{ss}$, $\Delta P_{0,syn}$, $\Delta P_{\infty,syn}$ ), including post-fault frequency and power dynamics. $\psi(\cdot)$ denotes the function of output metrics, $\varepsilon$ denotes the permissible error threshold. **d Schematic of the proposed constrained network algorithm.** The algorithm constructs a GCN incorporating physics-constrained knowledge, and employs combined mean squared error (MSE) and knowledge constraint (KC) loss functions for network training. The MSE loss function, a standard choice in supervised learning, quantifies the discrepancy between model predictions and actual values. The KC loss is a novel loss function introduced in this work. It penalizes violations of the predefined physical inequalities, ensuring the model's outputs remain consistent with the governing physical laws.

To achieve deep integration of knowledge and data for low-inertia power system pre-fault FSA with high accuracy, robustness, and generalization, we propose the GL-CN fusion algorithm. The GL algorithm directs the pre-training process based on physics-guided knowledge. In parallel, the CN algorithm structures the output layer via physics-constrained knowledge to enforce reliable and physically consistent predictions. The pseudo-code of the GL-CN algorithm is provided in **Supplementary Note 2**.

The GL algorithm is developed to integrate physics-guided knowledge into the DD framework, thus reducing dependence on massive datasets and ensuring model generalizability across diverse operating conditions. **Figure 2a** illustrates the schematic of physics-guided knowledge, which refers to the mapping between system frequency control parameters and post-fault security metrics (**Eq. 6** in **Methods**). Specifically, we obtain the analytical expression of post-fault security metrics by solving the ASFR model in time-domain (**Eqs. 7-10** in **Methods**). As an aggregated representation, the physics-guided knowledge captures relationships between system-level input and output. We therefore embed the physics-guided knowledge within the system-level data input channel (MLP channel) of the proposed framework by pre-training. Subsequently, the nodal data input channel (GCN channel) is trained using actual data. **Figure 2b** illustrates the GL algorithm, comprising the pre-training and fine-tuning stages, with the objective of incorporating universal physical principles before introducing actual operation data. In the pre-training stage, system parameters are sampled from a distribution with a broader range than that corresponding to typical operating conditions. The corresponding outputs are generated via physics-guided knowledge, creating a synthetic dataset. This process builds a physics-informed prior, enabling the model to learn generalized relationships and enhancing its generalization beyond the limited actual data distribution. Following pre-training, the model achieves accuracy comparable to that of KD methods while requiring substantially less training data during the subsequent fine-tuning stage. In this stage, the pre-trained model is further optimized using actual data to incorporate power flow and topological characteristics not captured during pre-training, therefore achieving accurate adaptation to the complex, nonlinear dynamics in real-world low-inertia power systems.

The CN algorithm is developed to integrate physics-constrained knowledge, ensuring the robustness of the fusion model output. **Figure 2c** illustrates the physics-constrained knowledge, which defines quantitative relationships among the six output metrics (**Eq. 15** in **Methods**). These relationships are derived from two governing principles of post-fault dynamics: (1) the post-fault frequency approaches its nadir along a near-parabolic trajectory (**Eq. 18** in **Methods**), and (2) the initial power imbalance is distributed according to unit inertia, whereas the quasi-steady-state imbalance is allocated based on unit regulation coefficients (**Eqs. 17, 21** in **Methods**). **Figure 2d** illustrates the CN algorithm, which constructs a graph where each node represents one of the six output security metrics. Edges are explicitly established between pairs of nodes whose corresponding metrics are governed by known physics-constrained relationships. Within this graph structure, connected nodes exchange information during processing, enforcing the physical constraints between the predicted outputs. The CN corrects output predictions by jointly optimizing both MSE loss and knowledge constraint (KC) loss during training (**Eq. S13** in **Supplementary Note 2**). During prediction, the model validates initial outputs against the KC loss. Predictions satisfying the KC threshold are retained, while those that violate it are corrected by the CN. This validation and correction process significantly enhances the robustness of the final model output.

## 2.3 Comparison of overall performance with existing methods

To validate the effectiveness of the knowledge-data fusion framework, we benchmarked the proposed methods against existing KD (ASFR) and DD (including support vector machine (SVM), k-nearest neighbors (KNN), convolutional neural network (CNN), and GCN) approaches under multiple scenarios. The hyperparameters of these models are detailed in **Methods**. In the case studies conducted prior to generalization testing, we utilized the full dataset of 12,707 samples spanning 97 operational scenarios (40–60% renewable penetration, provided in **Supplementary Note 2**). The samples were randomly partitioned into training (70%), validation (15%), and test (15%) sets. Model performance was evaluated using three metrics: maximum error (MAX), mean absolute percentage error (MAPE), and accuracy, formally defined in **Methods**.

**Figure 3a** presents the error distributions of all methods in scenarios with sufficient data and precise knowledge as violin plots, displaying the median values and interquartile ranges, and illustrating the impact of integrating data training, physics-guided, and physics-constrained knowledge. The results clearly demonstrate the effect of knowledge integration: in comparison with the baseline GCN, the knowledge-data fusion models exhibit substantially lower error distributions and significantly reduced median values. To further evaluate the framework, the testing is expanded to four scenarios based on data volume (sufficient/limited) and knowledge accuracy (precise/imprecise), as shown in **Fig. 3b**. For the scenarios with limited data, the training set was reduced from 8,876 to 951 samples. To simulate imprecise knowledge, parameters were adjusted to increase the knowledge label error rate from 10.1% to 39.0%. We evaluated the baseline GCN and our proposed methods under these conditions to delineate the specific contributions of each knowledge type. The results show that the GL method (incorporating physics-guided knowledge) performs superiorly in the data-limited scenarios, as the embedded physical principles effectively mitigate the issue of limited data. Conversely, the CN method (incorporating physics-constrained knowledge) demonstrates robust performance under imprecise knowledge, as its inherent correction mechanism mitigates the impact of label noise. The unified GL-CN method, incorporating both knowledge types, achieves optimal and robust performance across all tested scenarios.

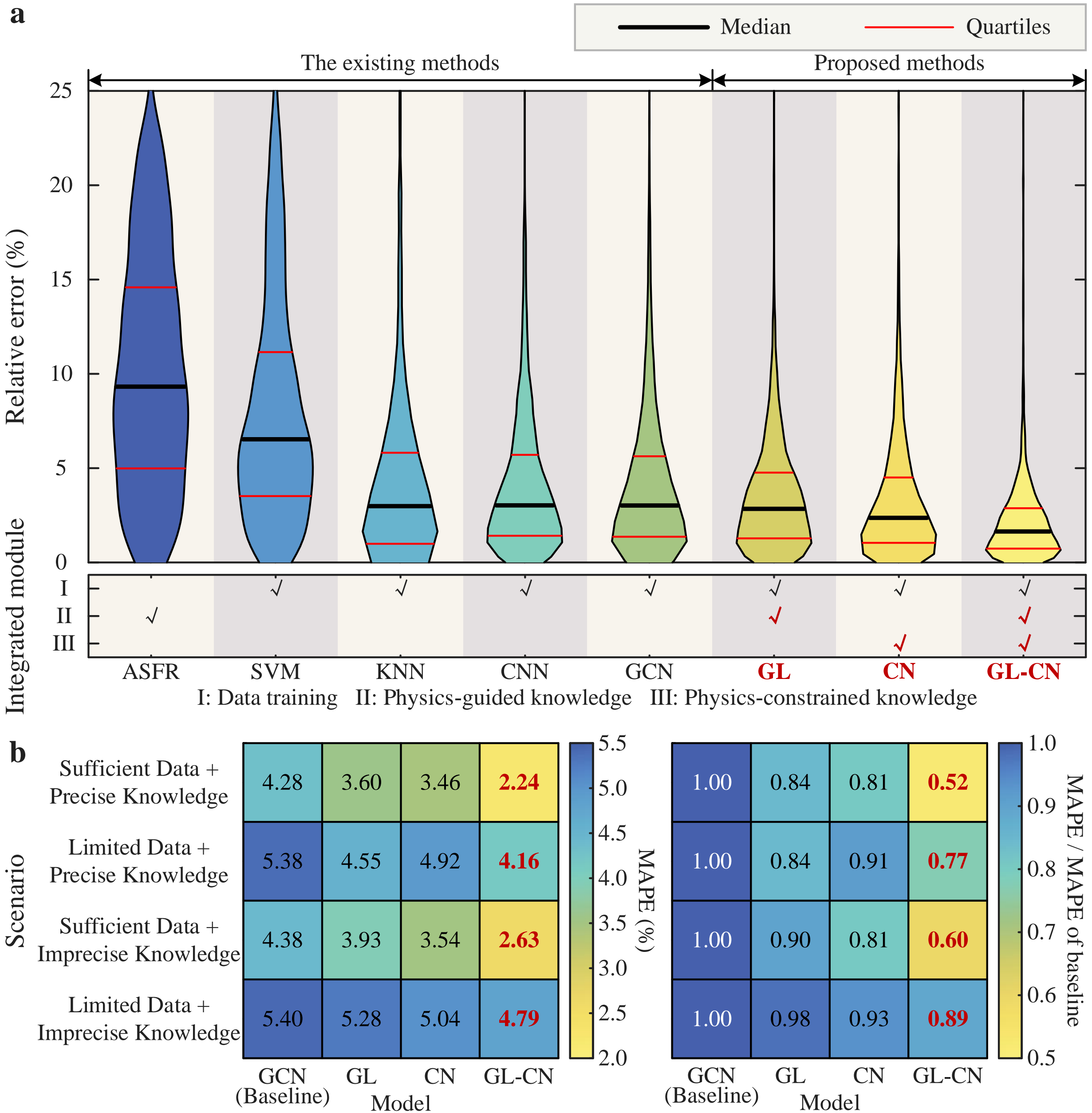

**Fig. 3: Comparison of the performance of the proposed methods against the existing methods.** **a Error distributions of the proposed and existing methods under scenarios with sufficient data and precise knowledge.** Violin plots depict the median and interquartile ranges. The tick symbols specify the use of data training, physics-guided knowledge, and physics-constrained knowledge for each method. **b Error heatmaps from ablation studies comparing the proposed methods against GCN benchmarks.** Performance is evaluated across four scenarios: sufficient/limited data with precise/imprecise knowledge. The left panel shows the MAPE for the proposed GL, CN, and GL-CN algorithms; the right panel shows the corresponding MAPE reduction relative to the GCN benchmarks.

## 2.4 Accuracy comparison of different methodologies

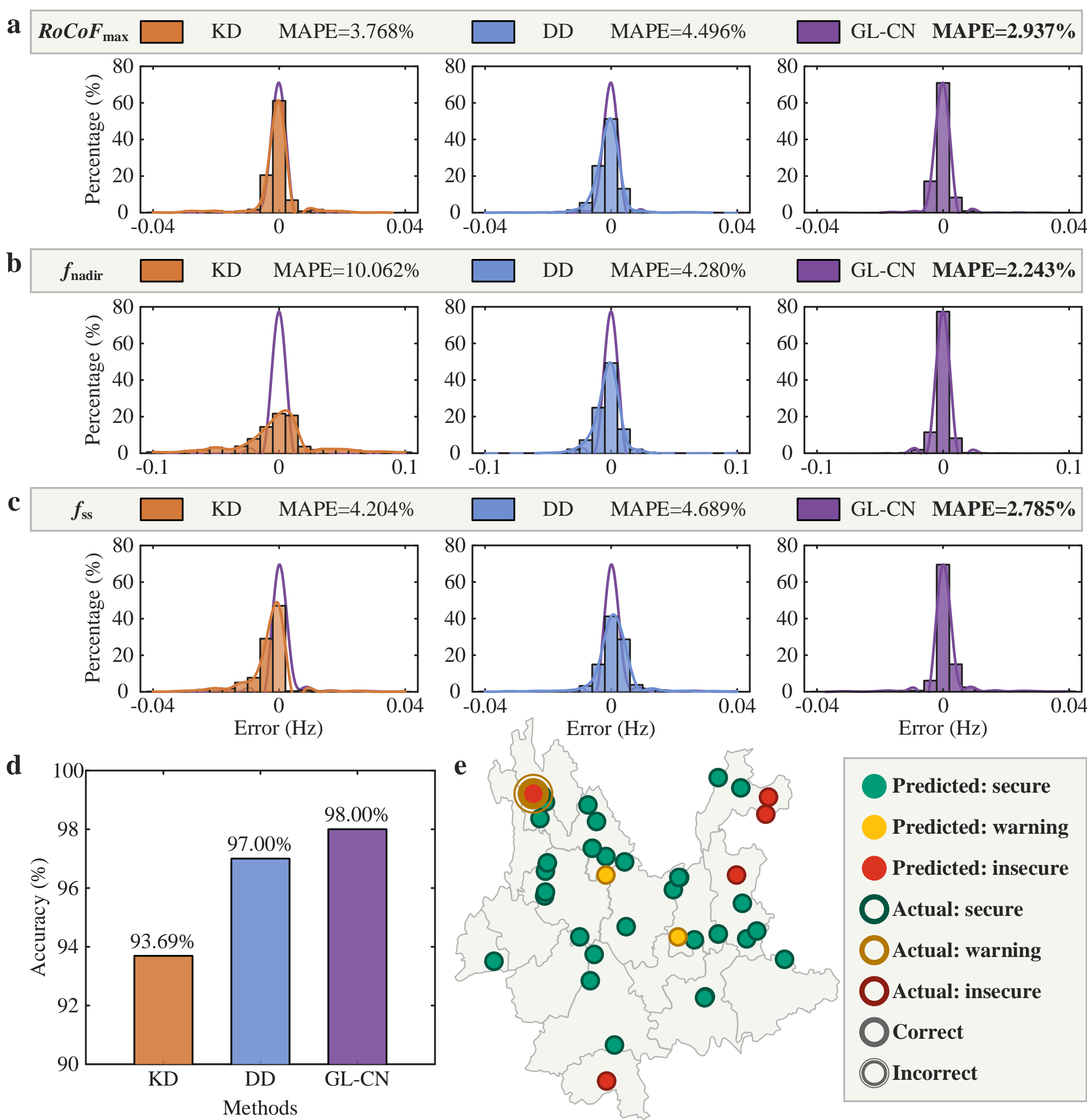

**Fig. 4: Performance comparison of the proposed method with KD and DD benchmarks with sufficient data and precise knowledge.**

**a, b, c Distribution of prediction errors for the $RoCoF_{max}$ (a), $f_{nadir}$ (b), $f_{ss}$ (c), comparing the proposed method with the KD (ASFR) and DD (GCN) benchmarks.** The histogram shows the error distribution of each method, and the curve represents the probability density. **d FSA accuracy of the proposed method and the KD, DD** **benchmarks. e FSA results of proposed method in winter operation mode of Yunnan Provincial Power Grid.** Each circle denotes an anticipated fault happening at the corresponding location. The fill color represents the FSA prediction from the proposed method, whereas the border color represents the actual FSA result. The border line style indicates the prediction accuracy for the corresponding fault.

Following a comprehensive evaluation of our proposed model's overall performance, a detailed analysis of its accuracy, robustness, and generalizability is provided in the subsequent sections. Specifically, in this section, we assess the model's accuracy via the design of comparative experiments against the state-of-the-art benchmark KD (ASFR) and DD (GCN) methods, under data-sufficient and knowledge-precise scenarios. This comparison analyzes both the distribution of prediction errors and the FSA accuracy to illustrate the distinct performance characteristics of each approach. **Figure 4a-c** depict the corresponding error distributions for the predictions of $RoCoF_{\mathrm{max}}$, $f_{\mathrm{nadir}}$, and $f_{\mathrm{ss}}$. For direct visual comparison, the error probability density curves are overlaid on the histograms to illustrate the performance advantage of the GL-CN framework. The figure reveals distinct error patterns across the methods: (1) The KD method provides straightforward calculations and high accuracy for metrics governed by a single dominant factor, such as the $RoCoF_{\mathrm{max}}$ and $f_{\mathrm{ss}}$. In contrast, for multi-factor metrics like $f_{\mathrm{nadir}}$, it yields significant prediction errors; (2) The DD method exhibits nonlinear mapping capabilities, maintaining comparable error distributions across all evaluated metrics; and (3) The proposed GL-CN algorithm displays sharper error distribution peaks than the DD method, which indicates a higher concentration of predictions within the low-error region and superior accuracy. **Figure 4d** compares the assessment accuracy of all three methods. While KD and DD achieve 93.69% and 97.00% accuracy, respectively, our GL-CN method reaches 98.00% accuracy. In summary, the knowledge-data fusion framework enhances prediction accuracy, providing system operators with an effective decision-support tool for pre-fault FSA and associated preventive control for low-inertia power systems.

After training, the model performs a two-step process: (1) predicting frequency responses for all anticipated faults under a given operating condition, and (2) classifying each predicted outcome against predefined frequency security thresholds (detailed in **Methods**) into three distinct security classes: secure, warning, and insecure. All results for each operating condition are aggregated and used to visualize frequency security risks. **Figure 4e** compares the frequency security risks under winter operating conditions, as assessed by TDS and the proposed method (The risk distributions of KD and DD benchmarks are provided in **Supplementary Note 3**). The predictions from the proposed method align closely with the simulation benchmarks, deviating only in a single case where an early-warning fault was misclassified as secure. Crucially, the proposed method achieves a substantial gain in computational efficiency, enabling rapid FSA. It evaluates all anticipated disturbances for a single operating condition in 0.7 seconds, compared to 1.5 hours for time-domain simulation.

## 2.5 Robustness to limited data and imprecise domain knowledge

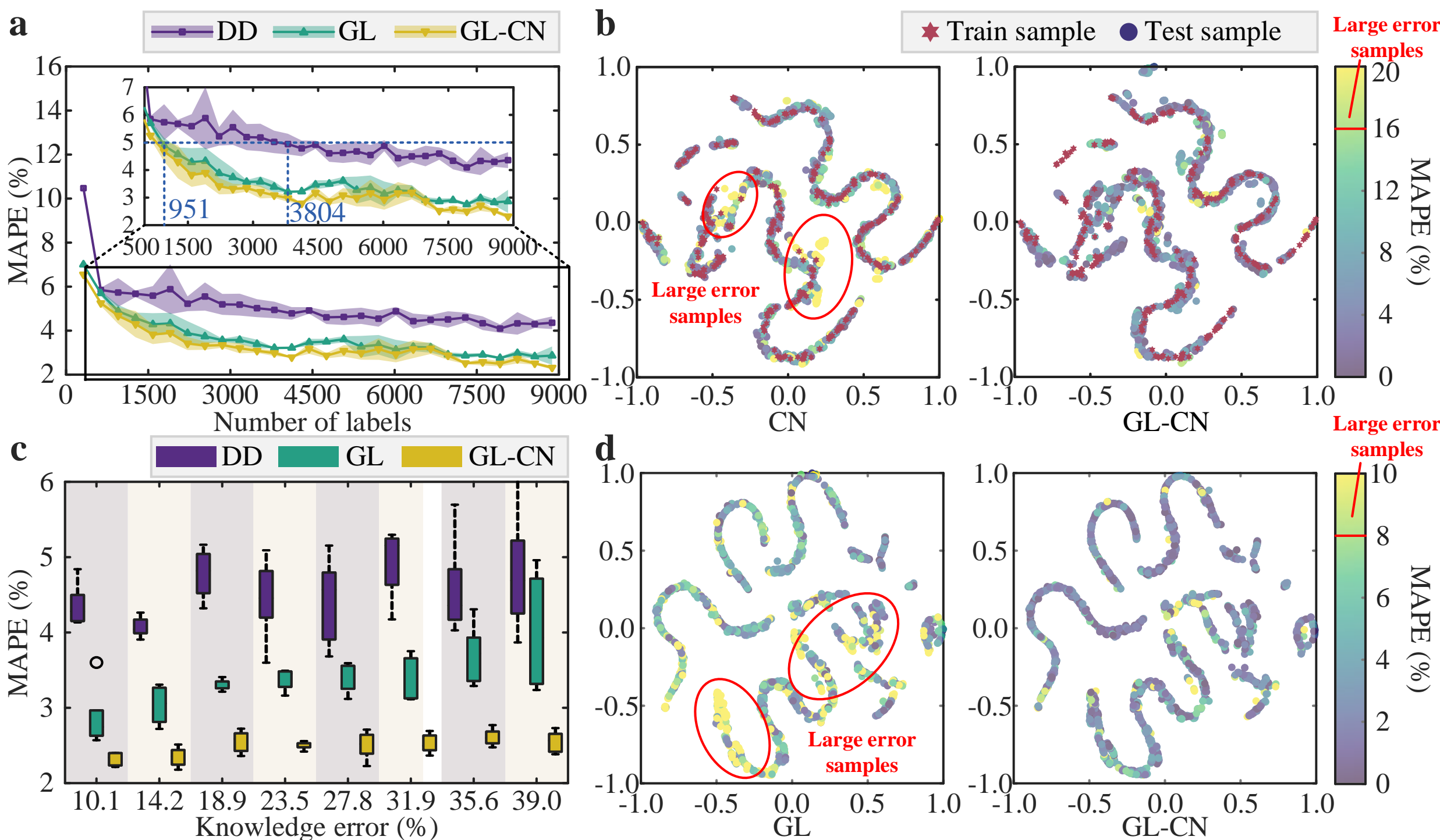

**Fig. 5: Robustness assessment of the GL and GL-CN models against the DD benchmark under scenarios with limited data or imprecise knowledge.**

**a MAPE comparison across training set size (317-8,876 samples) for all methods.** The inset panel shows a magnified view of the main figure, corresponding to the range of 500–9000 samples and a MAPE of 0–7%. **b t-SNE visualization of feature distributions (317 samples) comparing CN (left) and GL-CN (right).** Training samples are represented by red hexagon markers. Test samples (gradient circles) are colored by predicted MAPE values. **c Boxplots of MAPE under different knowledge errors (10.1%-39.0%) for all methods.** Increasing parameter errors induce corresponding prediction error growth in KD method. This figure reflects the sensitivity of the methods to knowledge accuracy. **d t-SNE feature distributions at 140% parameter error (39.02% knowledge error), comparing GL (left) and GL-CN (right).** Test samples (gradient circles) are colored by predicted MAPE values.

While knowledge-data fusion combines the strengths of both approaches, its accuracy remains dependent on both dataset size and knowledge precision. This dual dependency poses a critical challenge for low-inertia power system FSA applications. First, real-world frequency security events are scarce, requiring simulated data for training. However, in large-scale low-inertia power systems, each time-domain simulation takes about 40 seconds, making a 10,000-sample dataset prohibitively time-consuming (approximately 5 days). Furthermore, incomplete or uncertain system parameters introduces significant knowledge inaccuracies that can substantially degrade model performance. In summary, systematic evaluation of the knowledge-data fusion method's adaptability to varying training set sizes and knowledge precision thresholds is essential to establish its practical feasibility and minimum operation requirements.

The first experiment evaluates model performance across varying training sizes, ranging from 317 to 8,876 samples in 317-sample intervals, each condition repeated five times for statistical reliability. **Figure 5a** shows the MAPE dependence on training size for DD, GL, and GL-CN methods. All models show improvement in accuracy with increasing training samples until reaching a plateau. The error curve of GL-CN is slightly lower than that of GL, and both methods overall outperform DD, indicating that physics-guided knowledge integration in GL-based methods effectively compensates for limited data. With only 317 training samples, the DD method yields a MAPE of 10.47%, compared to 6.55% for GL-CN approach (achieve a 37.4% error reduction through knowledge integration). To achieve the target MAPE (<5%), our GL-CN method requires only 951 training samples (a 75% reduction compared to DD's 3,804-sample requirement). **Figure 5b** compares t-SNE visualizations of feature spaces learned by CN (left) and our GL-CN method (right) using only 317 labeled samples. Training samples are marked by red hexagon markers, while test samples appear as gradient circles with color intensity scaling from dark (0% MAPE) to light (20% MAPE). The DD method shows precise predictions (MAPE<10%) in cases covered by training samples but suffers catastrophic errors (MAPE>16%) in uncovered cases, revealing its limited generalization. In contrast, our GL-CN method maintains consistent performance (MAPE<16%) across most test samples, demonstrating that physics-guided knowledge fusion enhances generalization and reduces data requirements.

The second experiment evaluates model robustness to varying knowledge accuracy. We varied knowledge error from 10.1% to 39.0% by adjusting $T_R$ parameter errors in the physics-guided knowledge from 0% to 140%, which affect $f_{nadir}$ and $t_{nadir}$. The total knowledge error includes a fixed 10.1% from model simplification during aggregation, and a variable component (10.1% to 39.0%) from measurement inaccuracies in aggregation parameters, as detailed in the **Supplementary Note 4**. To assess the impact of parameter error, it was tested in 20% increments. For each condition, the model was independently trained five times for statistical robustness. **Figure 5c** displays the correlation between MAPE and knowledge accuracy for DD, GL, and GL-CN methods. Three distinct patterns emerge: (1) The error of the DD method fluctuates around 4.28%, unaffected by variations in knowledge accuracy; (2) GL performance degrades from 2.86% to 4.02% MAPE as knowledge errors increase; (3) The MAPE of the proposed GL-CN method exhibits only a marginal increase, rising from 2.32% to 2.55%. This robustness is a direct result of the physical constraints embedded via physics-constrained knowledge framework. Specifically, the CN detects and corrects inaccuracies in individual outputs, which preserves the overall accuracy of the model even in the presence of local knowledge errors. **Figure 5d** compares GL and GL-CN feature spaces under high parameter error (140%, 39.0% knowledge error) using t-SNE. Color gradient (dark to light, 0-10% MAPE) reveals GL-CN's superior robustness: while 10.8% of GL predictions exceed 8% MAPE and 35.8% below 2% MAPE, only 3.5% GL-CN predictions exceed 8%, with 51.9% remain below 2% error. This experiment confirms that the CN's physics-constrained architecture enhances model robustness to imprecise knowledge, maintaining reliable performance even with 39.0% knowledge error.

## 2.6 Generalization to untrained operating conditions

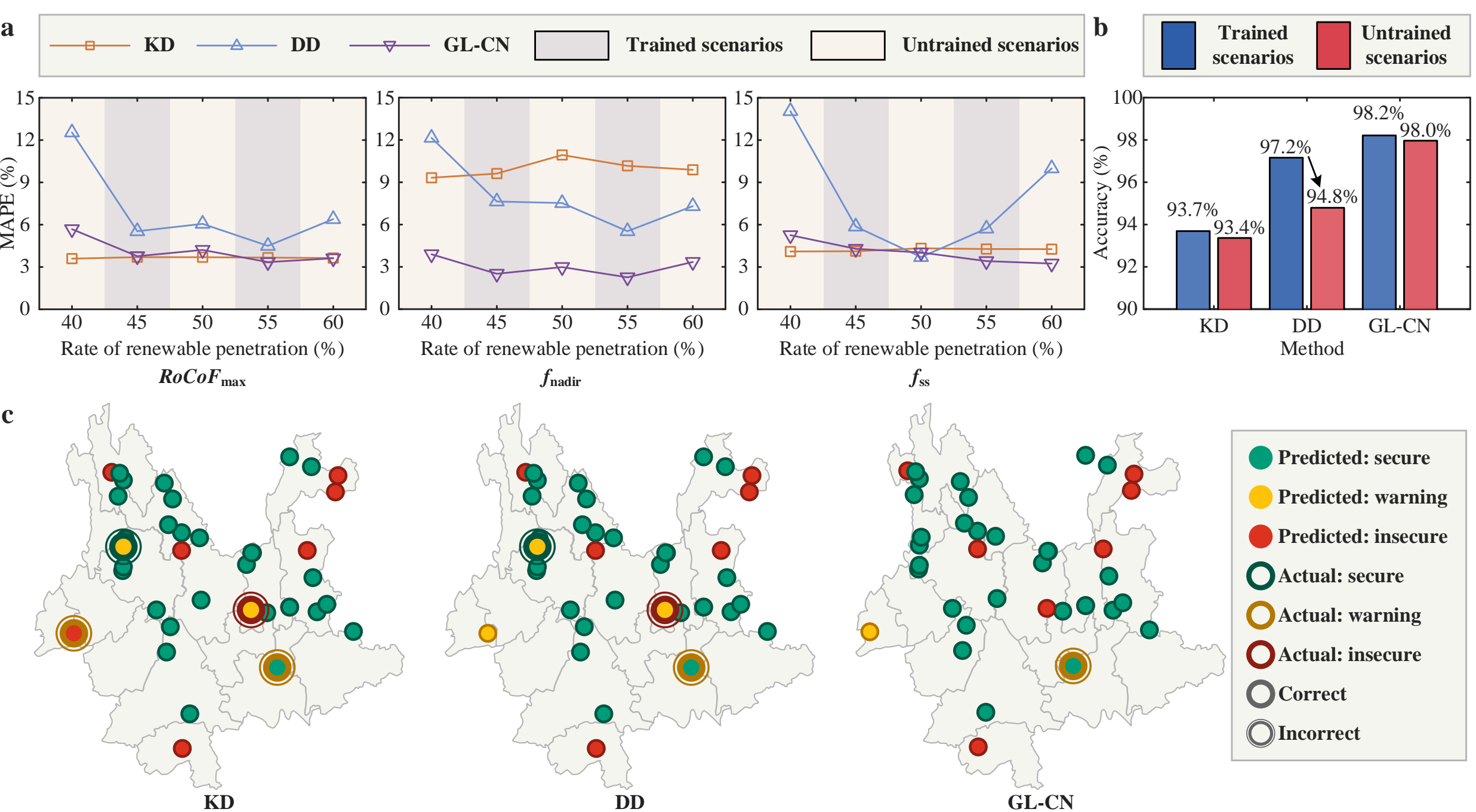

**Fig. 6: Generalization performance of the GL-CN method under untrained operating conditions, compared with KD and DD benchmarks.**

**a MAPE of the GL-CN method and the KD, DD benchmarks across varying renewable penetration levels.** Panels from left to right show results for $RoCoF_{\max}$, $f_{\text{nadir}}$ and $f_{\text{ss}}$. The shading distinguishes trained (dark gray) from untrained (light gray) scenarios. **b Bar charts of the FSA accuracy of the proposed methods and the KD, DD baselines in both trained and untrained scenarios. c FSA results of all methods in 60% renewable energy penetration scenario of Yunnan Provincial Power Grid. Panels from left to right show results of KD, DD and GL-CN methods.** Each circle denotes an anticipated fault occurring at the corresponding location. The fill color represents the FSA prediction from the proposed method, whereas the border color represents the actual FSA result. The border line style indicates the prediction accuracy for the corresponding fault.

Low-inertia power systems with high renewable penetration exhibit inherently complex and variable operating conditions. This is particularly evident in the Yunnan Provincial Power Grid, where seasonal hydropower availability drive significant shifts in system characteristics. Moreover, the growing renewable penetration means future power system grid conditions may not be captured in a model's original training data. Therefore, it is necessary to validate the generalization of the proposed knowledge-data fusion method under such untrained scenarios. To this end, the dataset from **Supplementary Note 2** was partitioned by renewable energy penetration rate. Samples at 45% and 55% penetration were split into training, validation, and test sets at a 70%:15%:15% ratio for model retraining. The remaining samples at 40%, 50%, and 60% penetration were held out as a generalization test set to evaluate performance under untrained scenarios in low-inertia power systems.

The generalization capability of the proposed GL-CN method is evaluated against KD and DD approaches by comparing the MAPE of its output predictions across varying renewable penetration levels, as shown in **Fig. 6a**. The KD method demonstrates strong generalization, as it is derived from general principles of power systems and is independent of training data. Provided accurate assessment of current $H$ and $R$ parameters, it maintains consistent prediction accuracy across varying renewable penetration levels. For simpler metrics such as the $RoCoF_{\mathrm{max}}$ and $f_{\mathrm{ss}}$, the KD method exhibits an error magnitude comparable to the GL-CN method. However, for the more complex $f_{\mathrm{nadir}}$, a metric governed by multiple factors, the MAPE of the KD method is substantially higher than the simple metrics due to the simplification of complex nonlinear factors. The DD model achieves low prediction errors at the trained penetration rates of 45% and 55%. However, its performance degrades catastrophically under untrained operating conditions. This fundamental lack of generalization poses a major barrier to the direct deployment of DD methods in real-world large-scale low-inertia power systems. The proposed GL-CN method integrates the strengths of both methods, achieving optimal performance across diverse renewable penetration levels. It exhibits superior generalization, maintaining high accuracy even at untrained penetration rates of 40%, 50%, and 60%.

To quantify generalization, the FSA accuracy of the three models in untrained scenarios was compared against their trained scenarios, as shown in **Fig. 6b**. The DD method shows a significant performance drop, with accuracy falling from 97% to 94%, confirming its poor generalization. In contrast, both the KD and GL-CN methods exhibit stable performance: the KD method consistently achieves 93% accuracy, while the GL-CN method sustains its superior accuracy of 98%, confirming its robust generalization capabilities. **Figure 6c** presents the FSA results of the GL-CN method and the KD, DD benchmark methods for a specific untrained scenario (60% renewable penetration). The GL-CN method maintains strong agreement with time-domain simulation benchmarks, further demonstrating its generalization capacity for power systems with variable operating conditions. More detailed comparison of errors and accuracy distribution is provided in **Supplementary Note 7**.

# 3 Discussion

Existing FSA methods depend largely on KD or DD models, which limits their accuracy and generalization. As a result, these methods are often inadequate for real-world low-inertia power systems, which are characterized by complex, dynamic, and diverse operating conditions. In this work, we propose a novel deep knowledge-data fusion framework that integrates low-inertia power system domain knowledge into both the architecture and the training process of deep learning models. This framework formalizes domain knowledge into two types (physics-guided and physics-constrained knowledge) and develops a corresponding fusion algorithm to enable their deep integration with deep learning models. During pre-training, the incorporation of physics-guided knowledge enables the model to learn generalized low-inertia power system frequency response mechanisms, enhancing its performance across diverse scenarios and reducing its reliance on extensive training data. During the prediction phase, the CN, trained based on physics-constrained knowledge, corrects the outputs to ensure their physical plausibility. The case study on Yunnan Provincial Power Grid in China demonstrated the framework's superiority in: (1) Accuracy: It achieves an average error of 2.24%, reducing the error by 77.7% and 47.7% compared to the KD and DD methods; (2) Robustness: It achieves under 5% average error with only 951 samples in data-limited scenarios, reducing sample needs by 75% against the DD methods, and retains under 3% average error despite 39.0% knowledge inaccuracies; (3) Generalization: It maintains 97.96% accuracy for FSA in unseen renewable penetration scenarios, achieving a 4.60% and 3.17% improvement over the KD and DD methods.

To the best of our knowledge, this work presents the first framework that deeply integrates multi-type power system knowledge with AI models by incorporating two main categories of domain knowledge, thereby significantly enhancing model accuracy, robustness, and generalization to meet industrial requirements. It can finish FSA of hundreds of steady-state operating conditions in just a few seconds while traditional simulation-based approaches may require days. In practice, low-inertia power systems encompass additional knowledge types, such as expert experience and operation guidelines expressed in natural language. The proposed framework is scalable to incorporate these knowledge types, potentially further enhancing accuracy and even interpretability if required by power industrial applications. Future work will focus on extending the representation and integration methods of heterogeneous knowledge types to further enhance the performance of the framework. Specifically, we will prioritize two directions: (1) Represention and integration of natural language knowledge[43-45]. Power systems possess abundant natural language knowledge, which often resists representations of mathematical equations, complicating its direct integration with AI models. Advances in large language model (LLM) technology facilitate the embedding of natural language knowledge into semantic vector spaces, allowing integration of natural language knowledge into AI models. (2) Knowledge extraction and reintegration[46-48]. The low-inertia power grid, as a complex nonlinear system, contains substantial knowledge that remains partially uncharacterized by humans, such as the impact of grid-following (GFL) and grid-forming (GFM) inverters[49-50] on post-fault frequency dynamics. We can utilize the strong nonlinear mapping capabilities of AI models to extract these unrepresented knowledge types from the data, thereby constructing a knowledge base. Then we reintegrate this knowledge base with the predictive model to enhance the performance. Furthermore, this knowledge extraction and reintegration strategy offers a pathway to break through conventional end-to-end black-box modeling by introducing interpretability in intermediate prediction processes.

While initially developed for low-inertia power system FSA, the proposed knowledge-data fusion framework exhibits strong potential for broader interdisciplinary applications. This broad applicability arises from the prevalence of both physics-guided and physics-constrained knowledge in engineering, where their integration with AI models can significantly improve overall performance. For instance, in weather forecasting, numerical methods are often limited by prohibitive computational expenses[51-52]. However, AI-based weather prediction struggles with accurately forecasting extreme events and ensuring physical consistency in its outputs[53-54]. Similar to power system FSA, weather forecasting relies on predictive models (Navier-Stokes equations), and its outputs must adhere to physical constraints such as the First Law of Thermodynamics and the ideal gas equation of state. Thus, meteorological knowledge can be analogously classified into two categories (i.e. physics-guided and physics-constrained knowledge) and fused with AI models via our framework to enhance robustness in extreme weather forecasting and strengthen the physical consistency of predictions. Furthermore, the proposed framework is applicable to diverse engineering problems (such as load forecasting[55], power market[56], battery state estimation[57] and wind speed forecasting[58]) exhibiting gray-box properties, enabling knowledge-data fusion. We believe our proposed framework has the potential to offer a promising approach to addressing complex engineering challenges in interdisciplinary domains.

# 4 Methods

## 4.1 Low-inertia power system FSA metrics

In steady-state operation, power system frequency is consistent across all nodes. During transient state, however, nodal frequencies can diverge slightly. To define a representative system-level frequency, the center of inertia frequency ($f_{\mathrm{COI}}$)[59] is therefore introduced. This metric constitutes a weighted average of individual nodal frequencies, where the weighting for each node corresponds to the inertia of its connected generators:

$$f_{\mathrm{COI}} = \frac{\sum_{i=1}^{N} H_i f_i}{\sum_{i=1}^{N} H_i} \tag{1}$$

where $N$ denotes the number of nodes in the power system. Specifically, $H_i$ is zero if no generator is connected to node $i$.

Four frequency security metrics, derived from $f_{\mathrm{COI}}$, are established to quantify post-fault frequency dynamic behavior. These metrics characterize frequency features at the inertia response, nadir, and quasi-steady-state stages. The overall frequency security status is determined by evaluating these metrics against pre-defined thresholds.

- **The inertia response stage**: The maximum rate of change of frequency ($RoCoF_{\max}$) quantifies the steepest instantaneous slope on the frequency trajectory after a fault. This extreme value typically occurs at the instant the fault occurs.

$$RoCoF_{\max} = \left.\frac{d\Delta f}{dt}\right|_{t=0^+} \tag{2}$$

where $\Delta f$ denotes the deviation of $f_{\mathrm{COI}}$. According to operation guidelines from the Yunnan Provincial Power Grid, a $RoCoF_{\max}$ exceeding 0.4 Hz/s triggers a warning state, and a $RoCoF_{\max}$ exceeding 0.5 Hz/s indicates an insecure state.

- **The nadir stage**: The nadir frequency ($f_{\mathrm{nadir}}$) is the maximum or minimum value on the post-fault frequency trajectory. A positive $\Delta P$, leading to frequency rise, produces the maximum frequency. Conversely, a negative $\Delta P$, causing frequency decline, results in the minimum frequency. The nadir time ($t_{\mathrm{nadir}}$) is defined as the instant $f_{\mathrm{nadir}}$ is reached.

$$f_{\mathrm{nadir}} = \begin{cases} \max\limits_{t}(\Delta f), \Delta P > 0 \\ \min\limits_{t}(\Delta f), \Delta P < 0 \end{cases} \tag{3}$$

$$t_{\mathrm{nadir}} = t\Big|_{\Delta f = f_{\mathrm{nadir}}} \tag{4}$$

where $\max(\cdot)$ denotes the maximum value of the curve, $\min(\cdot)$ denotes the minimum value of the curve. According to operation guidelines from the Yunnan Provincial Power Grid, a $f_{\mathrm{nadir}}$ exceeding 0.4 Hz triggers a warning state, and a $f_{\mathrm{nadir}}$ exceeding 0.5 Hz indicates an insecure state.

- **The quasi-steady-state stage**: The quasi-steady-state frequency ($f_{\mathrm{ss}}$) is the value the system frequency settles at following the inertia response and primary frequency control to a fault, prior to secondary control action.

$$f_{\mathrm{ss}} = \Delta f\Big|_{t=\infty} \tag{5}$$

According to operation guidelines from the Yunnan Provincial Power Grid, a $f_{\mathrm{ss}}$ exceeding 0.2 Hz triggers a warning state, and a $f_{\mathrm{ss}}$ exceeding 0.25 Hz indicates an insecure state.

## 4.2 Knowledge type I: physics-guided knowledge

Physics-guided knowledge is the knowledge type integrated via the guided learning algorithm. The physics-guided knowledge for FSA in low-inertia power systems is derived from the ASFR model. It is formally expressed as a mapping from power system frequency control parameters (inputs) to post-fault metrics (outputs):

$$(RoCoF_{\max}, \Delta P_{0,\text{syn}}, f_{\text{nadir}}, t_{\text{nadir}}, f_{\text{ss}}, \Delta P_{\infty,\text{syn}}) = \varphi(\Delta P, H, H_{\text{syn}}, R, R_{\text{syn}}, T_{\text{R}}) \tag{6}$$

where $\varphi(\cdot)$ denotes the mapping relationship from inputs to outputs.

Solving the ASFR model in the time domain provides the approximate analytical relationships between the frequency metrics and system-level parameters, A detailed analysis of error sources in these relationships is provided in **Supplementary Note 1**:

$$RoCoF_{\max} = \frac{1}{2H}\Delta P \tag{7}$$

$$t_{\text{nadir}} = \frac{1}{\omega_{\text{r}}}\arctan\left(\frac{T_{\text{R}}\omega_{\text{r}}}{T_{\text{R}}\zeta\omega_{\text{n}} - 1}\right) \tag{8}$$

$$f_{\text{nadir}} = \frac{\Delta P}{D + 1/R}\left(1 + be^{-\zeta\omega_{\text{n}}t_{\text{nadir}}}\right) \tag{9}$$

$$f_{\text{ss}} = \frac{1}{D + 1/R}\Delta P \tag{10}$$

The parameters $\omega_{\text{n}}$, $\zeta$, $\omega_{\text{r}}$, $b$ above are functions of power system parameters:

$$\omega_{\text{n}}^2 = \frac{D + 1/R}{2HT_{\text{R}}} \tag{11}$$

$$\zeta = \frac{DT_{\text{R}} + 2H + \alpha T_{\text{R}}/R}{2(D + 1/R)}\omega_{\text{n}} \tag{12}$$

$$\omega_{\text{r}} = \omega_{\text{n}}\sqrt{1 - \zeta^2} \tag{13}$$

$$b = \sqrt{1 - 2\zeta\omega_{\text{n}}T_{\text{R}} + \omega_{\text{n}}^2 T_{\text{R}}^2} \tag{14}$$

where $D$ denotes the system damping coefficient, which arises primarily from synchronous generators. Its value is often considered negligible. $\alpha$ is a weighting factor that aggregates the frequency response contributions from thermal, hydro, and renewable energy units.

## 4.3 Knowledge type II: physics-constrained knowledge

Physics-constrained knowledge is the knowledge type integrated via the constrained network algorithm. It consists of the distribution of the power imbalance and the trajectory of the frequency curve after faults. Together, they establish a set of constraint relationships among security metrics across different time periods:

$$\psi(RoCoF_{\max}, \Delta P_{0,\mathrm{syn}}, f_{\mathrm{nadir}}, t_{\mathrm{nadir}}, f_{\mathrm{ss}}, \Delta P_{\infty,\mathrm{syn}}) < \varepsilon \tag{15}$$

where $\psi(\cdot)$ denotes the function of output metrics, $\varepsilon$ denotes the permissible error threshold.

- **The inertia response stage**: In the initial inertia-response phase (0-1s post-fault), the power imbalance is distributed among generation units in proportion to their inertia. For modeling accuracy, the aggregated inertia used in this distribution accounts only for synchronous generators, as the effective inertia from renewable sources can be highly time-variant.

$$\Delta P_{0,i} = 2H_i RoCoF_{\max,i} \tag{16}$$

$$\Delta P_{0,\mathrm{syn}} = 2H_{\mathrm{syn}} RoCoF_{\max} \tag{17}$$

where $\Delta P_{0,i}$ denotes initial power imbalance allocation for the synchronous generator $i$ (inertia $H_i$). $\Delta P_{0,\mathrm{syn}}$ denotes initial power imbalance allocation for all the synchronous generators (inertia $H_{\mathrm{syn}}$).

- **The nadir stage**: An open-loop analysis of the frequency response in low-inertia systems (Ref. 31) shows that the trajectory preceding the frequency nadir approximates a parabolic curve. From the analytical expression of parabolic, approximate mathematical constraints among the frequency metrics can be derived.:

$$f_{\mathrm{nadir}} \approx \frac{1}{2} RoCoF_{\max} t_{\mathrm{nadir}} \tag{18}$$

- **The quasi-steady-state stage**: When the power system finishes primary frequency control and reaches quasi-steady-state (~60 s post-fault), the power imbalance is allocated among generation units based on their regulation coefficients. Consistent with the inertia response phase, this allocation is modeled using only synchronous generator parameters, as their governor response is well-defined compared to the variable regulation coefficients of renewable units.:

$$\Delta P_{\infty,i} = R_i f_{\mathrm{ss}} \tag{19}$$

$$\Delta P_{\infty,\mathrm{syn}} = R_{\mathrm{syn}} f_{\mathrm{ss}} \tag{20}$$

where $\Delta P_{\infty,i}$ denotes quasi-steady-state power imbalance allocation for the generator $i$ (regulation coefficients $R_i$), $\Delta P_{\infty,\mathrm{syn}}$ denotes quasi-steady-state power imbalance allocation for all the synchronous generators (regulation coefficients $R_{\mathrm{syn}}$).

## 4.4 Baseline methods

In the case studies, we employed several KD and DD models as benchmarks to evaluate the FSA performance of the proposed knowledge-data fusion model. The hyperparameters for all DD models were optimized to their best performance. The hyperparameters for all benchmark models are listed below:

- **ASFR:** $T_R$=10, Both $H$ and $R$ are obtained from power system parameter estimation.

- **SVM**: Radial basis function (RBF) kernel, penalty parameter $C$=1.0, $\gamma$ adaptively adjusted based on the size of dataset, $\varepsilon$=0.1.
- **KNN**: $k$=5, $p$=2, all neighbors have equal weight.
- **CNN**: It retains the dual-channel input structure but substitutes the GCN in the nodal input channel with a CNN. Furthermore, it lacks the CN module, using only the initial output channel. The CNN architecture comprises three layers, each with a 5×5 kernel, stride of 1, padding of 2, and is followed by a max-pooling layer.
- **GCN**: It retains the same input structure but lacks the two knowledge-integration components: the physics-guided pre-training (GL) and the physics-constrained output network (CN). Consequently, it operates as a DD model without embedded domain knowledge.

## 4.5 Model performance metrics

To evaluate the performance of the proposed framework and benchmark models comprehensively, three quantitative metrics are employed from distinct perspectives in this work:

- **MAX**: Defined as the largest absolute error observed across all test samples, this metric captures the model's worst-case prediction performance. It is computed as:

$$\mathrm{MAX} = \max_i(|y_i - \hat{y}_i|) \tag{21}$$

where $y_i$ denotes the true label of the sample $i$, $\hat{y}_i$ denotes the model prediction of the sample $i$.

- **MAPE**: Defined as the average relative prediction error across all test samples, representing the model's mean accuracy. It is computed as:

$$\mathrm{MAPE} = \frac{1}{N_{\mathrm{T}}}\sum_{i=1}^{N_{\mathrm{T}}}\left|\frac{y_i - \hat{y}_i}{y_i}\right| \tag{22}$$

where $N_{\mathrm{T}}$ denotes the number of samples in the test set.

- **Accuracy**: Defined as the proportion of correctly classified samples across the three categories (Secure, Warning, Unsecure), this metric evaluates the model's classification performance in application. It is computed as:

$$\mathrm{Accuracy} = \frac{1}{N_{\mathrm{T}}}(\mathrm{count}(y_i \in S_{\mathrm{S}} \wedge \hat{y}_i \in S_{\mathrm{S}}) + \mathrm{count}(y_i \in S_{\mathrm{W}} \wedge \hat{y}_i \in S_{\mathrm{W}}) + \mathrm{count}(y_i \in S_{\mathrm{I}} \wedge \hat{y}_i \in S_{\mathrm{I}})) \tag{23}$$

where $S_{\mathrm{S}}$, $S_{\mathrm{W}}$, $S_{\mathrm{I}}$ denotes the secure, warning, and unsecure sample sets, count(·) denotes the counting function.

## Data Availability

The data for model training and test is collected from the Yunnan Provincial Power Grid in China. Data required to generate figures in this paper are provided in our Code Ocean capsule, where non-publicly available data are shared in processed form, in accordance with nondisclosure requirements.

## Code Availability

The custom code used for implementing the deep learning models and generating the figures is available from the Code Ocean capsule.

## Acknowledgements

This research is supported by National Natural Science Foundation of China (U22B20111).

## Author contributions

Y.Z. contributed to the conceptualization, methodology, software, validation, and writing of the initial draft. W.Y. contributed to the conceptualization, reviewing, editing, supervision, and guidance. Y.L. contributed to methodology, formal analysis,visualization, and writing of the initial draft. H.S., W.G., S.W., and C.D. contributed to reviewing, editing, supervision, and guidance. J.W. and S.C. provided guidance to all co-authors throughout the process.

## Ethics declarations

### Competing interests